\documentclass[%
 reprint,
superscriptaddress,
 amsmath,amssymb,
 aps,
]{revtex4-2}

\usepackage{graphicx}
\usepackage{dcolumn}
\usepackage{bm}
\usepackage{mathrsfs}
\usepackage{float}

\begin{document}

\preprint{APS/123-QED}

\title{Prediction of the Optical Polarization and High Field Hyperfine Structure Via a Parametrized Crystal-Field Model for the Low Symmetry Centers in Er$^{3+}$ Doped Y$_{2}$SiO$_{5}$}

\author{N. L. Jobbitt}
\affiliation{School of Physical and Chemical Sciences, University of Canterbury, PB4800 Christchurch 8140, New Zealand}
\affiliation{The Dodd-Walls Centre for Photonic and Quantum Technologies, New Zealand}
\author{J.-P. R. Wells}
\email{jon-paul.wells@canterbury.ac.nz}
\author{M. F. Reid}
\email{mike.reid@canterbury.ac.nz}
\affiliation{School of Physical and Chemical Sciences, University of Canterbury, PB4800 Christchurch 8140, New Zealand}
\affiliation{The Dodd-Walls Centre for Photonic and Quantum Technologies, New Zealand}
\author{S. P. Horvath}
\affiliation{Department of Electrical Engineering, Princeton University, Princeton, NJ 08544, USA}
\author{P. Goldner}
\affiliation{Chimie ParisTech, PSL University, CNRS, Institut de Recherche de Chimie Paris, 75005 Paris, France}
\author{A. Ferrier}
 \affiliation{Chimie ParisTech, PSL University, CNRS, Institut de Recherche de Chimie Paris, 75005 Paris, France}
 \affiliation{Facult\'e des Sciences et Ing\'enierie, Sorbonne Universit\'e, UFR 933, 75005 Paris, France}

\date{\today}

\begin{abstract}
We report on the development and application of a parametrized crystal-field model for both C$_{1}$ symmetry centers in trivalent erbium-doped Y$_{2}$SiO$_{5}$. High resolution Zeeman and temperature dependent absorption spectroscopy was performed to acquire the necessary experimental data. The obtained data, in addition to the ground ($^{4}$I$_{15/2}$Z$_{1}$) state and exited ($^{4}$I$_{13/2}$Y$_{1}$) state Zeeman and hyperfine structure, was simultaneously fitted in order to refine an existing crystal-field interpretation of the Er$^{3+}$:Y$_{2}$SiO$_{5}$ system. We demonstrate that it is possible to account for the electronic, magnetic and hyperfine structure of the full 4f$^{11}$ configuration of Er$^{3+}$:Y$_{2}$SiO$_{5}$ and further, that it is possible to predict both optical polarization behavior and high magnetic field hyperfine structure of transitions in the 1.5 $\mu$m telecommunications band.
\end{abstract}

\maketitle


\section{\label{sec:intro}Introduction}

The last two decades have seen a global interest in the development of quantum information storage and communication devices in order to enhance the current classical computation and communication infrastructure \cite{cit:Nielsen}. Progress in this field has provided demonstrations of optical quantum memories, quantum gate implementations and single-photon sources \cite{Zhong1392, cit:de_riedmatten, Zhong, Ran_i__2017, PhysRevA.69.032307, PhysRevA.77.022307, cit:dibos}. Recently, demonstrations have been made showing control of multiple ions at the single-photon level \cite{Chen592}. Lanthanide-doped Y$_{2}$SiO$_{5}$ is an ideal material for the realization of such devices owing to the small nuclear moment of yttrium and low abundance of Si and O isotopes with non-zero nuclear spin \cite{PhysRevA.68.012320,FRAVAL2004347}. A direct consequence of this is the possibility of long storage times for information encoded into a qubit \cite{Wybourne}. This has enabled the observation of coherence times of over 1 min for Pr$^{3+}$:Y$_{2}$SiO$_{5}$; while Eu$^{3+}$:Y$_{2}$SiO$_{5}$ has exhibited a coherence time of over 6 h \cite{Zhong,PhysRevLett.111.033601}. The key technique used in order to obtain these coherence times is the zero-first-order-Zeeman (ZEFOZ) technique. This technique involves determining external magnetic field strengths at which the electronic structure of the system is insensitive to small fluctuations of the magnetic field in any direction. The field points at which this occurs are known as ZEFOZ points, which are avoided crossings of the hyperfine levels that exist in lanthanide-doped materials. At these points the dephasing induced by spin flips on neighboring host lattice ions is minimized, resulting in the long observed coherence times. As ZEFOZ points are located within the complex hyperfine structure of lanthanide-doped crystals, they have proven difficult to find experimentally, however, ZEFOZ points can be computationally predicted through the use of a spin Hamiltonian \cite{PhysRevB.66.035101}. The key advantage in using a spin Hamiltonian is that it more accurately models the magnetic and hyperfine structure of an individual state when compared to its crystal-field theory counterparts. When combined with a crystal-field model, which aims to reproduce the entire electronic configuration, a unique set of parameters can be obtained which then can be easily transferred between ions within the lanthanide series.

Trivalent erbium is of particular interest for quantum information applications as the ion possesses optical transitions in the well established 1.5 $\mu$m telecommunications band, the narrowest optical homogeneous linewidth observed to date (50 Hz) \cite{cit:SUN2002281}, in addition to a spin coherence time of over one second \cite{Ran_i__2017}. Furthermore, Er$^{3+}$ has a large hyperfine splitting relative to Pr$^{3+}$ and Eu$^{3+}$ \cite{horvath, PhysRevLett.111.033601,Zhong} which allows for larger memory bandwidths within these hyperfine transitions while still obtaining reasonably long coherence times.

Recently, crystal-field analyses have been performed for the C$_{1}$ symmetry centers in Ce$^{3+}$, Er$^{3+}$ and Yb$^{3+}$ doped Y$_{2}$SiO$_{5}$ \cite{cit:alizadeh, horvath,cit:Horvath_thesis, cit:zhou}; with further analyses under way for Y$_{2}$SiO$_{5}$ doped with Nd$^{3+}$, Sm$^{3+}$ and Ho$^{3+}$ \cite{Jobbitt2019, cit:jobbitt2, cit:mothkuri}.

We report on infrared to visible Zeeman absorption spectroscopy for both Er$^{3+}$ centers in Y$_{2}$SiO$_{5}$, the data from which culminates in a parametrized crystal-field model accounting for the energy level structure up to 27 000 cm$^{-1}$, approximately seventy electronic $g$ values measured along all three crystallographic axes as well as electron-paramagnetic resonance and Raman-heterodyne measurements obtained from the literature \cite{horvath, cit:Doualan_1995, cit:Sun, cit:Chen}. The directional data provided by our Zeeman measurements over tens of electronic levels, allows us to obtain a well determined, unique set of crystal-field parameters and therefore goes well beyond what has been reported previously. We demonstrate that, from this analysis, we can account for high magnetic-field hyperfine splittings as well as the optical polarization behavior in this technologically important material.

\section{\label{sec:experimental}Experimental}

Y$_{2}$SiO$_{5}$ is a monoclinic silicate crystal with C$^{6}_{2h}$ space group symmetry and lattice constants; $a = 10.4103$ \AA, $b = 6.7212$ \AA, $c = 12.4905$ \AA, and $ \beta   = 102^{\circ}39'$ \cite{cit:maksimov}. Here the crystallographic $b$ axis corresponds to the C$_{2}$ rotation axis and the crystallographic $a$ and $c$ axes are located in the mirror plane which is perpendicular to the crystallographic $b$ axis. Following the convention of Li \textit{et al.} we define the optical extinction axes as $D_{1}$ and $D_{2}$ which are located in the $a$-$b$ mirror plane and are perpendicular to each other in addition to the $b$ axis \cite{cit:li}. In this study we focus on the $X_{2}$ phase of Y$_{2}$SiO$_{5}$, which has two substitutional Y\textsuperscript{3+} sites, denoted as site 1 and site 2. Both of these sites have C$_{1}$ symmetry and here we follow the assignments made by \cite{cit:guillot}. Additionally, each site of Y$_{2}$SiO$_{5}$ also contains two magnetically inequivalent orientations, related by a 180$^{\circ}$ rotational symmetry, which arises from the C$_{2h}$ symmetry of the unit cell. This is particularly relevant in terms of Zeeman studies as the two orientations respond differently when a magnetic field is applied outside of the $D_{1}$-$D_{2}$ plane or the $b$ axis \cite{cit:Sun}.

The sample used in this study was grown in the $X_{2}$ phase of Y$_{2}$SiO$_{5}$ using the Czochralski process with an Er$^{3+}$ dopant concentration of 0.005 molar \%. The sample was grown to include natural abundances of erbium, of which $^{167}$Er is the only isotope to have a nuclear spin, with $I=7/2$. The crystal has dimensions of 6.27 mm along the $D_{1}$ axis, 6.18 mm along the $D_{2}$ axis, and 4.89 mm along the crystallographic $b$ axis.

High resolution temperature dependent and Zeeman spectroscopy was performed using a Bruker Vertex 80 Fourier transform infrared (FTIR) spectrometer, operated at a resolution of 0.1 cm$^{-1}$. Temperature dependent spectroscopy was performed by mounting the crystal on a copper holder and was cooled by thermal contact with a closed-cycle helium cryostat. The sample temperature was controlled by a temperature controller which adjusted the current through a resistive heater attached to the back of the sample cold finger. 

Zeeman spectroscopy was performed by attaching the sample to a copper mount which was then screwed into the bore of a 4 T Oxford Instruments superconducting solenoid built into a liquid helium cryostat. 

\section{\label{sec:results}Results and Discussion}

The 4f$^{11}$ configuration appropriate to Er$^{3+}$ has 182 doubly degenerate Kramers states, split into 41 multiplets. $^{167}$Er is the only naturally occurring isotope of erbium to have a non-zero nuclear spin, with $I = 7/2$.

\subsection{\label{sec:temp}Temperature dependent absorption spectroscopy}

Previously, the crystal-field electronic energy levels of both sites in Er$^{3+}$:Y$_{2}$SiO$_{5}$ have been measured, and reported, up to the $^{2}$H$_{11/2}$ multiplet at 20 000 cm$^{-1}$ \cite{cit:Doualan_1995}. We extend this work to 27 000 cm$^{-1}$ to include the $^{4}$F$_{7/2}$, $^{4}$F$_{5/2}$, $^{2}$H$_{9/2}$ and $^{4}$G$_{11/2}$ multiplets through the use of high resolution temperature dependent absorption spectroscopy.

\begin{figure*}
	\includegraphics[width=0.9\textwidth]{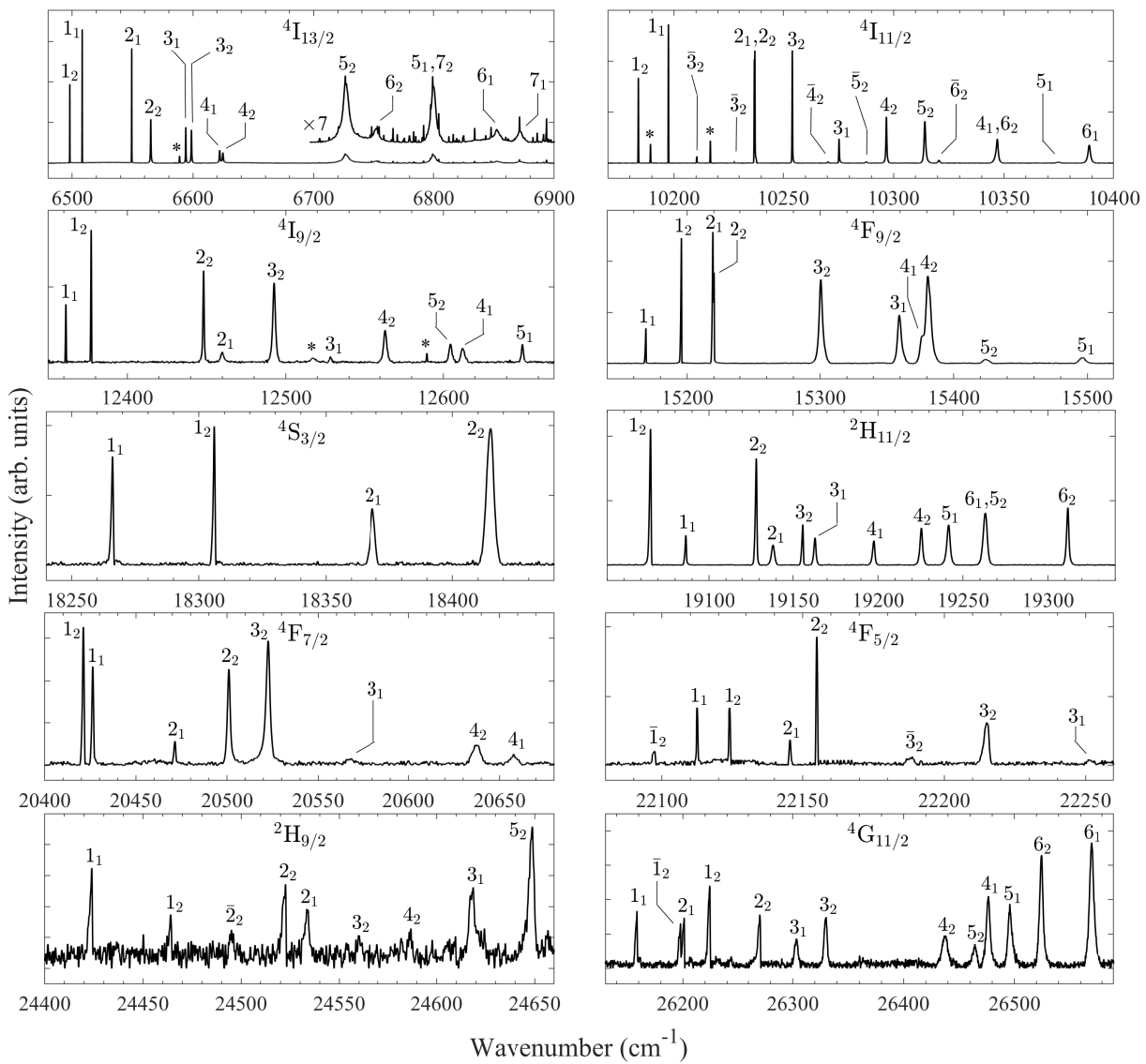}
	\caption{\label{fig:spectra}12 K absorption spectra of most excited multiplets up to the $^{4}$G$_{11/2}$ multiplet of Y$_{2}$SiO$_{5}$:0.005\%Er$^{3+}$. The transition labels indicate the ordering within the multiplet whilst a subscript gives the site assignment. Transitions from the Z$_{2}$ state are denoted with an overline. Spectral features labeled with an `*' are unrelated to Er$^{3+}$. Transitions to the $^{4}$F$_{3/2}$ multiplet could not be observed.}
\end{figure*}

Fig. \ref{fig:spectra} presents the 12 ~K absorption spectra of most excited multiplets up to the $^{4}$G$_{11/2}$ multiplet of Y$_{2}$SiO$_{5}$:0.005\%Er$^{3+}$ with all transitions assigned, with the exception of the $^{4}$F$_{3/2}$ multiplet and two levels of the $^{2}$H$_{9/2}$ multiplet for site 1 which were too weak to be observed. Levels were assigned to their respective sites by monitoring transitions from the excited states in the ground multiplet (hot lines) as the sample temperature was increased to 100 ~K. The 100 ~K spectra are omitted for brevity. The thermal population of states up to the $^{4}$I$_{15/2}$Z$_{4}$ state for both sites is evident at temperatures as low as 100 K. As the two sites of Er$^{3+}$:Y$_{2}$SiO$_{5}$ have an electronic energy level structure that are distinct from each other, each site therefore has a unique pattern which can be used to assign each spectral line found in absorption. Transitions from the $^{4}$I$_{15/2}$Z$_{1}$ state to their respective excited states are labeled as either belonging to site 1 or site 2 with a subscript. The spectral features labeled with an overbar are transitions from the $^{4}$I$_{15/2}$Z$_{2}$ state to the excited states. The extra structure seen in the 6 700 - 6 900 cm$^{-1}$ range of the $^{4}$I$_{13/2}$ multiplet is absorption due to residual atmospheric water vapor. Spectral features that could not be assigned as belonging to Er$^{3+}$ are marked with an `*'.

With the inclusion of this additional data, a total of 51 and 53 electronic states are assigned for site 1 and site 2 respectively. These values are summarized in Tables \ref{tab:site1} and \ref{tab:site2} respectively. With absorption to all multiplets remeasured here except the ground multiplet. The assignments made by Doualan \textit{et al.} were used for these states in the crystal-field fit \cite{cit:Doualan_1995}.

\subsection{\label{sec:zeeman}Zeeman absorption spectroscopy}

In order to achieve a definitive crystal-field analysis directional information is required. To achieve this, high resolution Zeeman absorption spectroscopy was performed for all three crystallographic axes, on the plethora of absorption lines observed through the infrared to the visible/near UV. Previous studies have determined the full $g$ tensors of the $^{4}$I$_{15/2}$Z$_{1}$ and $^{4}$I$_{13/2}$Y$_{1}$ states for both sites through rotational Zeeman and electron paramagnetic resonance experiments \cite{cit:Sun,cit:Chen}. 

\begin{figure}[h!]
\centering
	\includegraphics[width=0.5\textwidth]{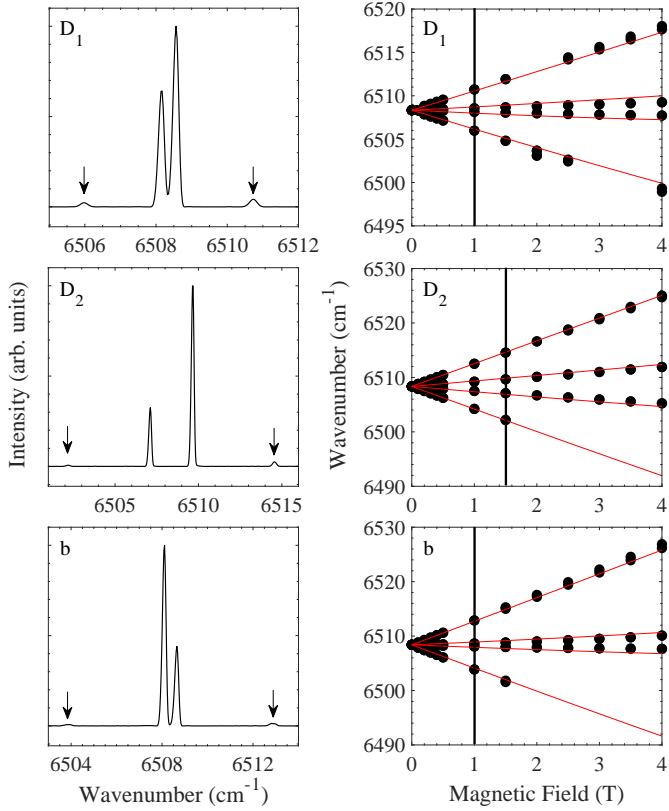}
	\caption{\label{fig:ErZeeman_site1}Magnetic splittings of the site 1 Z$_{1}$ $\rightarrow$ Y$_{1}$ transition for magnetic fields applied along the three crystallographic axes of Y$_{2}$SiO$_{5}$. The top, middle and bottom panels shows $B \parallel D_{1}$,  $B \parallel D_{2}$ and $B \parallel b$ respectively. The left panels show 4.2 K Zeeman absorption spectra at magnetic field strengths represented by the vertical lines in the right panels. The weak outer transitions are labeled with arrows to assist the reader. The right panels show the experimental splittings, represented by the circles, and the calculated splittings are represented by the red lines.}
\end{figure}

\begin{figure}[h!]
\centering
	\includegraphics[width=0.5\textwidth]{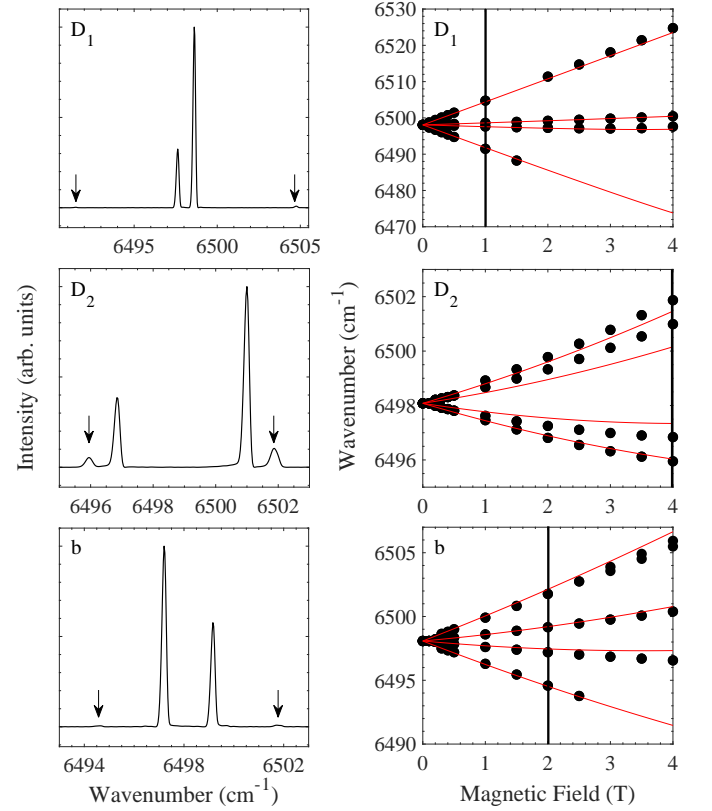}
	\caption{\label{fig:ErZeeman_site2}Magnetic splittings of the site 2 Z$_{1}$ $\rightarrow$ Y$_{1}$ transition for magnetic fields applied along the three crystallographic axes of Y$_{2}$SiO$_{5}$. The top, middle and bottom panels shows $B \parallel D_{1}$,  $B \parallel D_{2}$ and $B \parallel b$ respectively. The left panels show 4.2 K Zeeman absorption spectra at magnetic field strengths represented by the vertical lines in the right panels. The weak outer transitions are labeled with arrows to assist the reader. The right panels show the experimental splittings, represented by the circles, and the calculated splittings are represented by the red lines.}
\end{figure}

Figs. \ref{fig:ErZeeman_site1} and \ref{fig:ErZeeman_site2} show representative 4.2~K spectra of the $^{4}$I$_{15/2}$Z$_{1}$ $\rightarrow$ $^{4}$I$_{13/2}$Y$_{1}$ transition for site 1 and site 2 respectively under the influence of a magnetic field. The top, middle and bottom panels show the Zeeman splittings for magnetic fields directed parallel to the $D_{1}$, $D_{2}$ and $b$ axes respectively. The left panels show Zeeman absorption spectra at magnetic field strengths represented by the vertical lines in the right panels. The right panels show the experimental and calculated splittings as a function of magnetic field, with the calculated zero field energies shifted as appropriate to overlay the splittings. It can be seen that the calculations are a good approximation to the experimental data. Asymmetries in the spectra result from the quadratic Zeeman effect, due to repulsion by nearby states. 

A total of 68 $g$ values (20 along the $D_{1}$ axis, 23 along the $D_{2}$ axis and 25 along the $b$ axis) for site 1 and 70 $g$ values (25 along the $D_{1}$ axis, 24 along the $D_{2}$ axis and 21 along the $b$ axis) for site 2 were able to be determined. These values are summarized in Tables \ref{tab:site1} and \ref{tab:site2} respectively. The $g$ values could only be determined for states that have a splitting large enough to be resolved relative to the linewidth of the transition, within the linear splitting regime. For states above 22 000 cm$^{-1}$, most of the $g$ values could not be determined due to insufficient signal and relatively broad spectral lines.

\subsection{Parametrized crystal-field analysis}

The Hamiltonian appropriate for modeling the 4f$^{11}$ configuration of Er$^{3+}$ is given in Equation (\ref{eq:hamiltonian}). For more details the reader is directed to \cite{cit:liu}.
\begin{equation}
H = H_{FI} + H_{CF} + H_{Z} + H_{HF} + H_{Q}
\label{eq:hamiltonian}
\end{equation}
The terms in the equation correspond to the free-ion, crystal-field, Zeeman, the nuclear magnetic dipole hyperfine, and the nuclear quadrupole hyperfine interactions respectively. The free-ion interaction includes effects such as the configuration barycenter, parameterized by E$_\mathrm{avg}$, aspherical electrostatic repulsion, given by the Slater parameters, $F_{k}$, and the spin-orbit interaction, represented by $\zeta$, in addition to two- and three-body relativistic interactions as well as higher order effects. Here we fixed the $M^{0}$ and $P^{2}$ parameters, and constrained the $M^{2}$,  $M^{4}$, $P^{4}$ and $P^{6}$ parameters to the values and ratios defined by Ref. \cite{cit:carnall}. The crystal-field Hamiltonian used in this study has the form:
\begin{equation}
H_{CF} = \sum_{k,q}B^{k}_{q}C^{(k)}_{q}
\label{eq:CFhamiltonian}
\end{equation}
Here $k = 2, 4, 6$ and $q = -k,...,k$. The $B^{k}_{q}$ parameters are the crystal-field expansion coefficients and $C^{(k)}_{q}$ are spherical tensor operators using Wybourne’s normalization \cite{cit:Wybourne2}. All parameters with the exception of the axial ($q=0$) parameters are complex, leading to a total of 27 independent values for the C$_{1}$ point group symmetry appropriate to Y$_{2}$SiO$_{5}$. The Zeeman Hamiltonian has no free parameters while the magnetic dipole hyperfine, and the nuclear quadrupole interactions are represented by coupling constants $a_{l}$ and $a_{Q}$ respectively, which are required to be determined from experimental data \cite{cit:Horvath_thesis}.
In the fits reported here five free ion parameters, twenty seven crystal-field parameters and two hyperfine parameters were fitted to the experimental data. 

To obtain a unique fit for such a low symmetry requires orientational data. This is provided by measurements using magnetic fields in a variety of directions. 
The key magnetic-hyperfine data is reported using a spin Hamiltonian \cite{cit:Macfarlane}:
\begin{equation}
\mathscr{H} = \mu_{B} \mathbf{B} \cdot \mathbf{g} \cdot \mathbf{S} + \mathbf{I} \cdot \mathbf{A} \cdot \mathbf{S} + \mathbf{I} \cdot \mathbf{Q} \cdot \mathbf{I} - \mu_{n} g_{n} \mathbf{B} \cdot \mathbf{I}
\label{eq:spinhamiltonian}
\end{equation}
Here $\mu_{B}$ corresponds to the Bohr magneton, $\mathbf{B}$ is the magnetic field vector, $\mathbf{g}$, $\mathbf{A}$ and $\mathbf{Q}$ are the magnetic g, hyperfine and electric-quadrupole tensors respectively. $\mu_{n}$ is the nuclear magneton while $g_{n}$ is the nuclear g factor. $\mathbf{S}$ and $\mathbf{I}$ are vector representations of the electronic and nuclear spin operators respectively. 
Rather than fit to the tensors, in our fits we evaluate Hamiltonian (\ref{eq:hamiltonian}) at various magnetic field directions  \cite{horvath}. This results in a separate Hamiltonian matrix for each set of data, evaluated at a particular magnetic field sampled from the parametric spiral:
\begin{equation}
 \mathbf{B} = B_{0}
    \begin{bmatrix}
     \sqrt{1-t^2}\cos(6\pi t) \\
     \sqrt{1-t^2}\sin(6\pi t) \\
     t
    \end{bmatrix},
  \ \ \ t \in [-1,1]
\label{eq:B_vector}
\end{equation}
Here $B_{0}$ is the magnitude of the magnetic field vector. As multiple Hamiltonian matrices must be diagonalized simultaneously, the hyperfine portion of the fit used an intermediate coupled effective Hamiltonian in a basis spanning only the $^{4}$I$_{15/2}$ and $^{4}$I$_{13/2}$ multiplets. This reduced the dimension of the hyperfine Hamiltonian from 2912 to 224 states, reducing the time required to perform the fit. The matrix elements of the crystal-field levels were truncated to 40 000 cm$^{-1}$ to further increase performance. The starting parameters in the optimization routine were set to those found by Horvath \textit{et al.} \cite{horvath, cit:Horvath_thesis}. A coarse fit was performed using a basin hopping algorithm, which attempts a random step followed by a local minimization \cite{cit:wales, cit:Wales_2}. The Metropolis criterion was then applied to check if the random step is accepted and the algorithm was allowed to move to the newly found local minima \cite{cit:Metropolis}. The algorithm used for the local minimization was the bound optimization by quadratic approximation (BOBYQA) algorithm from the nonlinear-optimization (NLopt) package \cite{cit:johnson}. Following this, a final fit was performed using simulated annealing. Simulated annealing has the advantage that the parameter uncertainties can be estimated using Markov chain Monte Carlo (MCMC) techniques through sampling the posterior probability distribution \cite{cit:aster}.

For site 1 a total of 350 experimental data points were fitted simultaneously and are as follows:

\begin{itemize}
  \item 55 electronic energy levels up to the $^{4}$G$_{11/2}$ multiplet at $\sim$26 500 cm$^{-1}$ including the $^{4}$I$_{15/2}$Z$_{5}$ -- Z$_{8}$ states determined by Doualan \textit{et al.} \cite{cit:Doualan_1995}.
  \item 68 $g$ values (20  along  the  $D_{1}$ axis, 23 along the $D_{2}$ axis and 25 along the $b$ axis) corresponding to states up to the  $^{4}$F$_{5/2}$ multiplet at $\sim$20 500 cm$^{-1}$.
  \item 180 data points for the hyperfine splittings of the $^{4}$I$_{15/2}$Z$_{1}$ state, calculated from the $\mathbf{g}$, $\mathbf{A}$ and $\mathbf{Q}$ tensors determined by Chen \textit{et al.} \cite{cit:Chen}. The energy of each hyperfine state except the ground state (which was set as zero) was sampled at equally spaced intervals according to Equation (\ref{eq:B_vector}), with $B_{0} = 0.05$ T.
  \item 12 data points for the magnetic splittings of the $^{4}$I$_{13/2}$Y$_{1}$ state, calculated from the $\mathbf{g}$ tensor determined by Sun \textit{et al.} \cite{cit:Sun}, also sampled at equally spaced intervals according to Equation (\ref{eq:B_vector}), with $B_{0} = 0.05$ T.
  \item 4 data points from low-frequency Raman heterodyne measurements, calculated from a three dimensional curvature tensor of transition energy with respect to a magnetic field, obtained from \cite{horvath}. The four data points were sampled at magnetic fields of [0 0 0], [0.5 0 0], [0 0.5 0], [0 0 0.5] (with axes [$D_{1}$ $D_{2}$ $b$] and units of mT).
  \item 31 data points of high frequency Raman heterodyne spectroscopy along the $D_{2}$ direction as determined in \cite{horvath}. Second order polynomials were fitted and sampled at magnetic fields strengths of 0.0, 0.3 and 0.5 mT. Nine transitions were fitted using this method whereas four of the transitions have a very steep gradient around zero field and therefore were only sampled at 0.5 mT.
\end{itemize}

To date there has been no Raman heterodyne studies performed on site 2 of Er$^{3+}$:Y$_{2}$SiO$_{5}$ and therefore 319 experimental data points were fitted simultaneously and are as follows:

\begin{itemize}
  \item 57 electronic energy levels up to the $^{4}$G$_{11/2}$ multiplet at $\sim$26 500 cm$^{-1}$ including the $^{4}$I$_{15/2}$Z$_{5}$ -- Z$_{8}$ states determined by Doualan \textit{et al.} \cite{cit:Doualan_1995}.
  \item 70 $g$ values (25  along  the  $D_{1}$ axis, 24 along the $D_{2}$ axis and 21 along the $b$ axis) corresponding to states up to the  $^{4}$F$_{5/2}$ multiplet at $\sim$20 500 cm$^{-1}$.
  \item 180 data points for the hyperfine splittings of the $^{4}$I$_{15/2}$Z$_{1}$ state, calculated from the $\mathbf{g}$, $\mathbf{A}$ and $\mathbf{Q}$ tensors determined by Chen \textit{et al.} \cite{cit:Chen}. The energy of each hyperfine state except the ground state (which was set as zero) was sampled at equally spaced intervals according to Equation (\ref{eq:B_vector}), with $B_{0} = 0.05$ T.
  \item 12 data points for the magnetic splittings of the $^{4}$I$_{13/2}$Y$_{1}$ state, calculated from the $\mathbf{g}$ tensor determined by Sun \textit{et al.} \cite{cit:Sun}, also sampled at equally spaced intervals according to Equation (\ref{eq:B_vector}), with $B_{0} = 0.05$ T.
\end{itemize}

\begin{table*}
    \caption{Calculated and experimental electronic energies levels and $g$ values up to 27 000 cm$^{-1}$ for site 1 in Er$^{3+}$:Y$_{2}$SiO$_{5}$. All energies are in cm$^{-1}$. Levels marked with a `--' were not assigned. Levels marked with an `*' are assignments made by Doualan \textit{et al.} \cite{cit:Doualan_1995}.} 
    \begin{ruledtabular}
    \begin{tabular}{ cccccccccccccc }
     & & \multicolumn{3}{c}{Energies}  & &  \multicolumn{8}{c}{$g$ values} \\\cline{3-5}\cline{7-14}
 & & & &  & &  \multicolumn{2}{c}{$D_{1}$ axis}  & & \multicolumn{2}{c}{$D_{2}$ axis} & & \multicolumn{2}{c}{$b$ axis} \\\cline{7-8}\cline{10-11}\cline{13-14}
        Multiplet & State & Calc. & Exp. & Difference  && Calc. & Exp. && Calc. & Exp. && Calc. & Exp.  \\\hline
        $^{4}$I$_{15/2}$& Z$_{1}$ & -10   &  0    & 10  &&  5.47 & 5.46  && 11.06  & 10.91 && 8.18  & 8.36   \\
						& Z$_{2}$ & 39    & 39    & 0   &&  6.67 &  --   &&   1.50 & --    && 8.19  &   --   \\
						& Z$_{3}$ & 72    & 84    & 12  &&  3.98 &  --   &&  7.07  & --    && 2.35  &     -- \\
		                & Z$_{4}$ & 111   & 102   & -9  &&  3.93 &  --   &&  9.34  & --    && 4.23  &     -- \\
						& Z$_{5}$ & 179   & 172*  & -7  &&  4.08 &  --   &&  8.58  & --    && 7.35  &     -- \\
						& Z$_{6}$ & 416   & 424*  & 8   && 10.22 &  --   &&  7.62  & --    && 3.68  &     -- \\
						& Z$_{7}$ & 474   & 481*  & 7   && 10.79 &  --   &&  7.14  & --    &&  4.42 &     -- \\
		                & Z$_{8}$ & 509   & 513*  & 4   &&  8.30 &  --   && 12.08  & --    && 6.50  &   --   \\
		$^{4}$I$_{13/2}$& Y$_{1}$ & 6518  & 6508  & -10 && 3.93  & 4.64  && 6.82   &  6.90 && 10.24 & 10.00  \\
						& Y$_{2}$ & 6564  & 6549  & -15 && 5.93  & 4.24  && 4.35   & 4.41  && 5.69  &  7.77  \\
						& Y$_{3}$ & 6599  & 6598  & -1  && 6.59  & 4.54  && 4.73   &  3.04 && 2.56  &  6.10  \\
						& Y$_{4}$ & 6637  & 6624  & -13 && 3.43  & 9.32  && 6.29   &  5.44 && 5.97  &  1.87  \\
		                & Y$_{5}$ & 6805  & 6800  & -5  && 8.47  &  --   && 6.16   & --    && 2.43  &  --    \\
						& Y$_{6}$ & 6866  & 6852  & -14 && 9.13  &  --   && 5.57   & --    && 4.56  &  --    \\
						& Y$_{7}$ & 6887  & 6871  & -16 && 7.32  &  --   && 9.66   & --    && 5.23  & --     \\
		$^{4}$I$_{11/2}$& A$_{1}$ & 10189 & 10197 & 8   && 3.30  & 3.89  && 3.67   &  4.80 && 8.75  &   8.23 \\
						& A$_{2}$ & 10231 & 10237 & 6   && 4.43  &  3.31 && 3.15   &  --   &&  4.04 &  5.99  \\
						& A$_{3}$ & 10258 & 10275 & 17  && 4.76  &  5.41 && 2.84   & 4.00  &&  4.27 &  3.30  \\
		                & A$_{4}$ & 10345 & 10347 & 2   && 6.58  & 3.51  && 4.44   & --    && 1.82  &   --   \\
						& A$_{5}$ & 10386 & 10375 & -11 && 6.87  &  5.33 && 4.13   &  5.31 && 3.73  &   --   \\
						& A$_{6}$ & 10396 & 10389 & -7  && 5.65  &  7.28 && 7.11   &  6.21 && 3.79  &   1.66 \\
	    $^{4}$I$_{9/2}$ & B$_{1}$ & 12337 & 12362 & 25  && 4.39  &  3.75 && 3.16   &  3.68 && 1.64  &   1.87 \\
						& B$_{2}$ & 12430 & 12461 & 31  && 1.99  &  --   && 1.40   &  2.10 && 5.51  &   4.03 \\
						& B$_{3}$ & 12532 & 12529 & -3  && 1.61  &  --   && 3.08   &  2.01 && 3.27  &   3.73 \\
						& B$_{4}$ & 12606 & 12613 & 7   && 2.60  &  --   && 3.47   &  --   && 1.55  &   1.65 \\
    		            & B$_{5}$ & 12656 & 12651 & -5  && 3.31  &  3.66 && 3.27   &  1.98 && 2.26  &   2.44 \\
		$^{4}$F$_{9/2}$	& D$_{1}$ & 15185 & 15170 & -15 && 8.60  &  7.86 && 0.68   & 1.76  && 3.42  &  4.61  \\
						& D$_{2}$ & 15236 & 15220 & -16 && 2.68  &  1.39 && 7.51   &  6.84 && 2.05  &  3.21  \\
		                & D$_{3}$ & 15360 & 15360 & 0   && 2.54  &  0.54 && 2.66   & --    && 2.72  &   2.30 \\
						& D$_{4}$ & 15402 & 15377 & -25 && 2.41  &   --  && 4.90   & --    && 2.67  &     -- \\
						& D$_{5}$ & 15493 & 15497 & 4   && 4.04  &   --  && 3.64   & --    && 5.88  &     -- \\
		$^{4}$S$_{3/2}$ & E$_{1}$ & 18272 & 18267 & -5  && 4.13  &  3.64 && 2.03   &  3.00 && 2.84  &   2.77 \\
						& E$_{2}$ & 18359 & 18369 & 10  && 2.43  &   --  && 2.44   &  2.24 && 4.37  &     -- \\
		$^{2}$H$_{11/2}$& F$_{1}$ & 19115 & 19087 & -28 && 2.66  & 8.47  && 9.21   &  7.93 && 4.89  &   0.89 \\
						& F$_{2}$ & 19152 & 19139 & -13 && 5.97  &   --  && 6.51   & --    && 2.62  &     -- \\
						& F$_{3}$ & 19177 & 19164 & -13 && 4.12  &   --  && 5.45   &  2.08 && 7.23  & 10.36  \\
						& F$_{4}$ & 19242 & 19242 & 0   && 0.38  &   --  && 2.60   &  4.72 && 3.67  &   3.79 \\
						& F$_{5}$ & 19277 & 19264 & -13 && 3.88  &   --  && 7.64   & --    && 0.68  &     -- \\
						& F$_{6}$ & 19302 & 19312 & 10  && 4.64  &  4.93 && 3.52   &  4.80 && 5.15  &   5.20 \\
		$^{4}$F$_{7/2}$	& G$_{1}$ & 20424 & 20430 & 6   && 4.77  &  5.91 && 3.99   &  3.92 && 3.16  &   3.25 \\
						& G$_{2}$ & 20468 & 20475 & 7   && 2.91  &   --  &&  3.84  &  4.58 && 3.16  &  1.90  \\
						& G$_{3}$ & 20557 & 20570 & 13  && 3.79  &   --  &&  2.79  & --    && 2.56  &     -- \\
						& G$_{4}$ & 20633 & 20660 & 27  && 3.00  &   --  &&  2.43  & --    && 6.97  &     -- \\
		$^{4}$F$_{5/2}$	& H$_{1}$ & 22099 & 22115 & 16  && 1.29  &   --  &&  1.64  & --    && 2.08  &   1.18 \\
						& H$_{2}$ & 22130 & 22148 & 18  && 3.40  &   --  &&  1.96  & --    && 1.51  &     -- \\
						& H$_{3}$ & 22217 & 22255 & 38  && 1.18  &   --  &&  2.76  & --    && 2.54  &     -- \\
		$^{4}$F$_{3/2}$	& I$_{1}$ & 22470 & --    & --  && 1.90  &   --  &&  1.08  & --    && 0.72  &     -- \\
						& I$_{2}$ & 22601 & --    & --  && 0.80  &   --  &&  0.85  & --    && 1.36  &     -- \\
		$^{2}$H$_{9/2}$ & K$_{1}$ & 24403 & 24423 & 20  && 5.70  &   --  &&  3.49  & --    && 2.49  &     -- \\
						& K$_{2}$ & 24515 & 24534 & 19  && 2.18  &   --  &&  1.79  & --    && 5.53  &     -- \\
						& K$_{3}$ & 24594 & 24618 & 24  && 2.08  &   --  &&  2.41  & --    && 4.51  &     -- \\
						& K$_{4}$ & 24639 & --    & --  && 2.18  &   --  &&  3.77  & --    && 3.69  &     -- \\
						& K$_{5}$ & 24686 & --    & --  && 3.88  &   --  &&  3.78  & --    && 2.66  &     -- \\
		$^{4}$G$_{11/2}$& L$_{1}$ & 26207 & 26159 & -48 && 4.51  &   --  &&  8.68  & --    && 5.03  &     -- \\
						& L$_{2}$ & 26223 & 26202 & -21 && 6.93  &   --  &&  5.44  & --    && 4.47  &     -- \\
						& L$_{3}$ & 26322 & 26304 & -18 && 4.74  &   --  &&  3.95  & --    && 2.72  &     -- \\
						& L$_{4}$ & 26469 & 26477 & 8   && 1.88  &   --  &&  3.85  & --    && 4.04  &     -- \\
						& L$_{5}$ & 26514 & 26497 & -17 && 4.58  &   --  &&  6.93  & --    && 1.13  &     -- \\
						& L$_{6}$ & 26577 & 26571 & -6  && 3.82  &   --  &&  5.95  & --    && 3.43  &     -- \\

    \end{tabular}
    \end{ruledtabular}
    \label{tab:site1}
\end{table*}

\begin{table*}
    \caption{Calculated and experimental electronic energies levels and $g$ values up to 27 000 cm$^{-1}$ for site 2 in Er$^{3+}$:Y$_{2}$SiO$_{5}$. All energies are in cm$^{-1}$. Levels marked with a `--' were not assigned.  Levels marked with an `*' are assignments made by Doualan \textit{et al.} \cite{cit:Doualan_1995}.} 
    \begin{ruledtabular}
    \begin{tabular}{ cccccccccccccc }
     & & \multicolumn{3}{c}{Energies}  & &  \multicolumn{8}{c}{$g$ values} \\\cline{3-5}\cline{7-14}
 & & & &  & &  \multicolumn{2}{c}{$D_{1}$ axis}  & & \multicolumn{2}{c}{$D_{2}$ axis} & & \multicolumn{2}{c}{$b$ axis} \\\cline{7-8}\cline{10-11}\cline{13-14}
        Multiplet & State & Calc. & Exp. & Difference  && Calc. & Exp. && Calc. & Exp. && Calc. & Exp.  \\\hline
        $^{4}$I$_{15/2}$& Z$_{1}$ & -1    & 0     & 1   && 14.65 & 15.31 && 2.21   &  2.66 && 3.13  &   2.84 \\
						& Z$_{2}$ & 29    & 27    & -2  &&  8.58 &  --   &&  5.27  & --    && 4.93  &   --   \\
						& Z$_{3}$ & 66    & 62    & -4  &&  3.07 &  --   &&  6.40  & --    && 6.95  &     -- \\
		                & Z$_{4}$ & 129   & 126   & -3  &&  5.21 &  --   && 11.48  & --    && 4.50  &     -- \\
						& Z$_{5}$ & 170   & 169*  & -1  &&  6.52 &  --   &&  4.03  & --    && 6.98  &     -- \\
						& Z$_{6}$ & 313   & 314*  & 1   &&  3.71 &  --   &&  7.17  & --    && 7.18  &     -- \\
						& Z$_{7}$ & 347   & 350*  & 3   &&  6.57 &  --   && 11.73  & --    && 3.72  &     -- \\
		                & Z$_{8}$ & 408   & 415*  & 7   &&  5.18 &  --   &&  4.11  & --    && 13.96 &   --   \\
		$^{4}$I$_{13/2}$& Y$_{1}$ & 6515  & 6498  & -17 && 12.37 & 13.14 && 0.70   &  0.46 && 5.01  & 4.93   \\
						& Y$_{2}$ & 6570  & 6565  & -5  && 7.82  & 8.23  && 3.61   & 2.16  && 2.46  &  5.45  \\
						& Y$_{3}$ & 6596  & 6594  & -2  && 3.36  & 2.61  && 9.33   & 10.75 && 4.04  &  2.63  \\
						& Y$_{4}$ & 6637  & 6622  & -15 && 5.73  & 6.04  && 3.35   &  4.75 && 5.72  &  4.36  \\
		                & Y$_{5}$ & 6733  & 6726  & -7  && 3.15  &  --   && 4.85   & --    && 7.24  &  --    \\
						& Y$_{6}$ & 6767  & 6752  & -15 && 6.11  &  --   && 10.58  & 10.88 && 3.04  &  --    \\
						& Y$_{7}$ & 6813  & 6800  & -13 && 3.88  &  --   && 3.60   & --    && 11.37 & --     \\
		$^{4}$I$_{11/2}$& A$_{1}$ & 10176 & 10184 & 8   && 9.60  & 10.04 && 0.30   & --    && 4.07  &   --   \\
						& A$_{2}$ & 10223 & 10237 & 14  && 5.82  &  6.71 && 3.80   & 5.19  &&  2.81 &  3.92  \\
						& A$_{3}$ & 10243 & 10254 & 11  && 3.36  & 3.06  && 2.66   & 5.44  &&  5.02 &  --    \\
		                & A$_{4}$ & 10296 & 10297 & 1   && 2.80  & 3.91  && 3.78   & 3.05  && 4.71  &   3.71 \\
						& A$_{5}$ & 10317 & 10314 & -3  && 4.34  &  5.53 && 8.12   &  5.95 && 1.49  &   --   \\
						& A$_{6}$ & 10347 & 10347 & 0   && 2.74  &  1.14 && 2.88   &  4.42 && 8.95  &  9.35  \\
	    $^{4}$I$_{9/2}$ & B$_{1}$ & 12357 & 12378 & 21  && 0.62  &  1.17 && 4.17   &  3.08 && 3.00  &   4.02 \\
						& B$_{2}$ & 12424 & 12449 & 25  && 4.23  &  3.23 && 2.24   &  2.05 && 2.87  &   1.05 \\
						& B$_{3}$ & 12471 & 12494 & 23  && 1.85  &  3.54 && 2.52   &  --   && 2.61  &   --   \\
						& B$_{4}$ & 12542 & 12564 & 22  && 2.18  &  2.00 && 3.67   & 3.63  && 2.54  &  3.53  \\
    		            & B$_{5}$ & 12599 & 12605 & 6   && 2.27  &  --   && 1.76   &  --   && 4.39  &   --   \\
		$^{4}$F$_{9/2}$	& D$_{1}$ & 15211 & 15196 & -15 && 2.79  & 3.72  && 6.26   &  5.83 && 3.89  &   5.02 \\
						& D$_{2}$ & 15234 & 15221 & -13 && 2.90  & 2.40  && 3.00   &  --   &&  3.35 &  5.20  \\
		                & D$_{3}$ & 15309 & 15302 & -7  && 3.90  & 4.41  && 3.03   & 4.11  && 4.44  &  3.73  \\
						& D$_{4}$ & 15390 & 15382 & -8  && 6.66  &   --  && 1.99   & --    && 2.68  &     -- \\
						& D$_{5}$ & 15437 & 15425 & -12 && 7.66  &   --  && 3.76   & --    && 2.17  &     -- \\
		$^{4}$S$_{3/2}$ & E$_{1}$ & 18307 & 18306 & -1  && 1.91  & 1.83  && 2.66   &  2.99 && 4.14  &   7.22 \\
						& E$_{2}$ & 18405 & 18415 & 10  && 4.77  &   --  && 1.30   &  --   && 1.91  &     -- \\
		$^{2}$H$_{11/2}$& F$_{1}$ & 19093 & 19067 & -26 && 6.29  & 7.39  && 8.74   &  8.77 && 3.89  &   3.42 \\
						& F$_{2}$ & 19143 & 19129 & -14 && 7.92  & 8.47  && 2.68   & 4.57  && 3.91  &  2.75  \\
						& F$_{3}$ & 19166 & 19157 & -9  && 0.76  & 2.36  && 5.58   &  4.72 && 8.98  &  8.04  \\
						& F$_{4}$ & 19203 & 19199 & -4  && 2.24  & 2.16  && 4.47   &  3.97 && 3.89  &   4.41 \\
						& F$_{5}$ & 19242 & 19226 & -16 && 3.65  &   --  && 7.81   & 8.00  && 3.04  &     -- \\
						& F$_{6}$ & 19255 & 19264 & 9   && 5.70  &   --  && 3.82   &  --   && 4.86  &   --   \\
		$^{4}$F$_{7/2}$	& G$_{1}$ & 20421 & 20424 & 3   && 2.19  &  2.74 && 3.08   &  3.57 && 6.85  &   6.47 \\
						& G$_{2}$ & 20489 & 20504 & 15  && 1.44  &   --  &&  3.16  &  5.01 && 3.61  &  --    \\
						& G$_{3}$ & 20519 & 20526 & 7   && 4.55  &   --  &&  3.22  & --    && 2.11  &     -- \\
						& G$_{4}$ & 20623 & 20640 & 17  && 7.30  &   --  &&  0.95  & --    && 2.90  &     -- \\
		$^{4}$F$_{5/2}$	& H$_{1}$ & 22109 & 22126 & 17  && 0.76  &  0.62 &&  1.55  & --    && 3.79  &     -- \\
						& H$_{2}$ & 22146 & 22157 & 11  && 2.39  &   --  &&  3.24  & --    && 1.85  &   1.35 \\
						& H$_{3}$ & 22196 & 22218 & 22  && 3.98  &   --  &&  0.89  & --    && 1.31  &     -- \\
		$^{4}$F$_{3/2}$	& I$_{1}$ & 22454 & --    & --  && 0.74  &   --  &&  1.36  & --    && 1.59  &     -- \\
						& I$_{2}$ & 22584 & --    & --  && 1.54  &   --  &&  0.22  & --    && 0.71  &     -- \\
		$^{2}$H$_{9/2}$ & K$_{1}$ & 24442 & 24464 & 22  && 1.40  &   --  &&  5.75  & --    && 3.23  &     -- \\
						& K$_{2}$ & 24513 & 24522 &  9  && 3.28  &   --  &&  3.05  & --    && 3.11  &     -- \\
						& K$_{3}$ & 24556 & 24557 & 1   && 3.42  &   --  &&  2.73  & --    && 3.28  &     -- \\
						& K$_{4}$ & 24605 & 24587 & -18 && 1.99  &   --  &&  4.23  & --    && 3.14  &     -- \\
						& K$_{5}$ & 24657 & 24649 & -8  && 3.22  &   --  &&  3.62  & --    && 3.87  &     -- \\
		$^{4}$G$_{11/2}$& L$_{1}$ & 26251 & 26225 & -26 && 6.43  &   --  &&  9.77  & --    && 3.78  &     -- \\
						& L$_{2}$ & 26278 & 26271 & -7  && 5.74  &   --  &&  2.34  & --    && 8.61  &     -- \\
						& L$_{3}$ & 26335 & 26331 & -4  && 3.36  &   --  &&  2.77  & --    && 5.10  &     -- \\
						& L$_{4}$ & 26435 & 26438 & 3   && 3.29  &   --  &&  4.59  & --    && 3.57  &     -- \\
						& L$_{5}$ & 26485 & 26465 & -20 && 3.46  &   --  &&  8.73  & --    && 2.76  &     -- \\
						& L$_{6}$ & 26519 & 26525 & 6   && 6.70  &   --  &&  4.91  & --    && 2.70  &     -- \\

    \end{tabular}
    \end{ruledtabular}
    \label{tab:site2}
\end{table*}
\begin{table*}
	\caption{Fitted values for the free-ion, crystal-field and hyperfine parameters and their related uncertainties of site 1 and site 2 in Er$^{3+}$:Y$_{2}$SiO$_{5}$. All values are in cm$^{-1}$. Parameters determined by Horvath \textit{et al.} are also included for comparison \cite{horvath, cit:Horvath_thesis}.} 
	\begin{ruledtabular}
	\begin{tabular}{ ccccccccc }
	\rule{0pt}{10pt}	          & \multicolumn{4}{c}{Site 1} && \multicolumn{3}{c}{Site 2} \\\cline{2-5}\cline{7-9} 
	\rule{0pt}{10pt}   Parameter  & This study & Uncertainty & Ref. \cite{horvath} & Ref. \cite{cit:Horvath_thesis} &&  This study & Uncertainty & Ref. \cite{cit:Horvath_thesis} \\\hline
	\rule{0pt}{10pt} $E_\mathrm{avg}$  & 35491.3 & 0.1 &  35503.5 & --  && 35507.5  & 0.1 & -- \\
    $F^{2}$       & 95805.7          & 1.0          & 96029.6         & 95346     && 96121.9       & 1.3        & 95721\\
	$F^{4}$       & 67869.7          & 3.4          & 67670.6         & 68525     && 67722.4       & 4.5        & 68564\\
	$F^{6}$       & 53148.2          & 2.5          & 53167.1         & 52804     && 53241.2       & 3.1        & 52999\\
	$\zeta$       & 2360.5           & 0.1          & 2362.9          & 2358      && 2362.3        & 0.1        &  2356\\
	$B^{2}_{0}$   & -479.6           & 6.1          & -149.8          & -563      && 389.0         & 3.7        &  354\\
	$B^{2}_{1}$   & 471.4+143.8i     & 2.9 + 3.0i   & 420.6+396.0i    & 558+280i  && -325.7-95.8i  & 2.7 + 3.0i & 498.6807+274i\\
	$B^{2}_{2}$   & 125.5-2.0i       & 2.8 + 2.3i   & -228.5+27.6i    & 143-121i  && -368.5+53.7i  & 1.8 + 2.0i & -75.8028+60i\\
	$B^{4}_{0}$   & -640.6           & 31.3         & 1131.2          & -125      && 17.2          & 15.5       & 226\\
	$B^{4}_{1}$   & 288.8+924.1i     & 7.2 + 25.3i  & 985.7+34.2i     & 225-831i  && -378.7-519.5i & 5.1 + 9.3i & -657.8381+593i\\
	$B^{4}_{2}$   & -273.9+320.9i    & 11.1 + 16.7i & 296.8+145.0i    & -48-945i  && -72.0-146.0i  & 5.7 + 6.7i & 335.7827+253i\\
	$B^{4}_{3}$   & -873.7-367.8i    & 20.7 + 9.7i  & -402.3-381.7i   & -615-688i && -890.8+570.4i & 9.5 + 7.3i & -71.3262-46i\\
	$B^{4}_{4}$   & -600.8+1210.5i   & 23.7 + 9.2i  & -282.3+1114.3i  & 744-102i  && -198.7-567.9i & 7.8 + 5.2i & -813.9654+64i\\
	$B^{6}_{0}$   &  145.7           & 13.2         & -263.2          & -28       && 73.4          & 4.3        & 219\\
	$B^{6}_{1}$   & -105.9-329.0i    & 2.9 + 4.0i   & 111.9+222.9i    & 49+199i   && -37.5+49.9i   & 3.4 + 5.7i & -127+197i\\
	$B^{6}_{2}$   & -119.9+164.1i    & 7.7 + 8.8i   & 124.7+195.9i    & 120-107i  && 135.5+60.6i   & 4.5 + 1.5i & -36-47i\\
	$B^{6}_{3}$   &  1.1+133.3i      & 6.7 + 4.5i   & -97.9+139.7i    & 195-55i   && -166.7+131.8i & 2.6 + 4.0i & 17-108i\\
	$B^{6}_{4}$   & -84.6+36.9i      & 5.0 + 4.5i   & -93.7-145.0i    & -287-161i && 227.2+47.6i   & 1.2 + 3.0i & -100+77i\\
	$B^{6}_{5}$   &  75.5+6.9i       & 4.3 + 6.6i   & 13.9+109.5i     & -117+162i && 119.5+64.3i   & 3.7 + 3.2i & -263+103i\\
	$B^{6}_{6}$   & -48.5+118.0i     & 6.2 + 4.2i   & 3.0-108.6i      & 136+186i  && 37.6-41.3i    & 3.5 + 2.8i & 12-26i\\
    $S^{2}$       & 386.6            &   --         & 399.0           & 483.0     && 363.1         &   --       & 397.9\\
	$S^{4}$       & 948.2            &    --        & 862.9           & 824.6     && 653.3         &   --       & 607.5\\
	$S^{6}$       & 183.8            &   --         & 189.6           & 218.6     && 151.5         &   --       & 171.4\\
	$a_{l}$       & 0.005306         & 0.000008     & 0.005466        & 0.0059    && 0.005389      & 0.000012   & 0.0069\\
	$a_{Q}$       & 0.0554           & 0.0020       & 0.0716          & 0.0800    && 0.0240        &  0.0024    & 0.0808\\
	\end{tabular}
	\end{ruledtabular}
	\label{tab:Er_Parameters}
\end{table*}
\begin{table}
	\caption{Parameters that were held fixed during the fitting process and were set to those found by Carnall \textit{et al.} in Er$^{3+}$:LaF$_{3}$ \cite{cit:carnall}.} 
	\begin{ruledtabular}
	\begin{tabular}{ cc }
	Parameter     & Value (cm$^{-1}$) \\\hline
	$\alpha$      & 17.79\\
	$\beta$       & -582.10\\
	$\gamma$      & 1800.00\\
	$T^{2}$       & 400.00\\
	$T^{3}$       & 43.00\\
	$T^{4}$       & 73.00\\
	$T^{6}$       & -271.00\\
	$T^{7}$       & 308.00\\
	$T^{8}$       & 299.00\\
	$M^{0}$     & 3.86\\
	$P^{2}$     & 594.00\\
	\end{tabular}
	\end{ruledtabular}
	\label{tab:Fixed_Parameters}
\end{table}
\begin{table*}
	\caption{Spin Hamiltonian parameters for the $^{4}$I$_{15/2}$Z$_{1}$ and $^{4}$I$_{13/2}$Y$_{1}$ states for site 1 and site 2 in Er$^{3+}$:Y$_{2}$SiO$_{5}$. All $A$ and $Q$ values are in MHz.} 
	\begin{ruledtabular}
	\begin{tabular}{ clccccccccccccc }
	\rule{0pt}{10pt}	         & & \multicolumn{6}{c}{Site 1} && \multicolumn{6}{c}{Site 2} \\\cline{3-8}\cline{10-15} 
	\rule{0pt}{10pt} State  & Study & $g_{xx}$ & $g_{yy}$ & $g_{zz}$ & $g_{xy}$ & $g_{xz}$ & $g_{yz}$ &&  $g_{xx}$ & $g_{yy}$ & $g_{zz}$ & $g_{xy}$ & $g_{xz}$ & $g_{yz}$ \\\hline
	$^{4}$I$_{15/2}$Z$_{1}$ & This study      & 3.03     & 8.75     & 4.92     & -3.00    & -3.44    & 5.74     && 14.47    & 1.51     & 1.65     & -1.89    & 2.37     & 0.02  \\
	                        & Ref. \cite{horvath}  & 2.10     & 8.37     & 5.49     & -3.43    & -3.21    & 5.16     &&  --      & --       & --       & --       & --       & -- \\
	                        & Ref. \cite{cit:Chen} & 2.90     & 8.90     & 5.12     & -2.95    & -3.56    & 5.57     && 14.37    & 1.93     & 1.44     & -1.77    & 2.40     & -0.43 \\
	                        & Ref. \cite{cit:Sun}  & 3.07     & 8.16     & 5.79     & -3.14    & -3.40     & 5.76    && 14.65    & 1.97     & 0.90     & -2.12    & 2.55     & -0.55 \\
	$^{4}$I$_{13/2}$Y$_{1}$ & This study      & 1.63     & 3.66     & 8.29     & -1.86    & -2.98    & 5.11     && 11.58    & 0.31     & 2.13     & -0.69    & 4.44     & -0.26 \\
	                        & Ref. \cite{horvath}  & 2.04     & 4.44     & 7.94     & -2.24    & -3.40    & 5.15     && --       & --       & --       & --       & --       & -- \\
	                        & Ref. \cite{cit:Sun}  & 1.95     & 4.23     & 7.89     & -2.21    & -3.58     & 4.99    && 12.03    & 0.21     & 1.77     & -5.85    & 4.52     & -0.30 \\
                        	&                 & $A_{xx}$ & $A_{yy}$ & $A_{zz}$ & $A_{xy}$ & $A_{xz}$ & $A_{yz}$ && $A_{xx}$ & $A_{yy}$ & $A_{zz}$ & $A_{xy}$ & $A_{xz}$ & $A_{yz}$ \\\hline
	$^{4}$I$_{15/2}$Z$_{1}$ & This study      & 317.20   & 917.44   & 515.83   & -314.96  & -361.63  & 602.05   && 1541.77  & 159.17   & 175.52   & -200.91  & 253.34   & 1.61  \\
                        	& Ref. \cite{horvath}  & 200.80   & 911.27   & 586.95   & -344.23  & -362.61  & 586.95   &&  --      & --       & --       & --       & --       & -- \\
                        	& Ref. \cite{cit:Chen} & 274.29   & 827.50   & 706.15   & -202.52  & -350.82  & 635.15   && -1565.3  & -15.3    & 127.8    & 219.0    & -124.4   & -0.7 \\
    $^{4}$I$_{13/2}$Y$_{1}$ & This study      & 206.60   & 466.60   & 1061.93  & -236.41  & -380.64  & 653.32   &&  1504.47 & 40.73    & 276.53   & -90.45   & 577.51   & -34.26 \\
                        	& Ref. \cite{horvath}  & 271.96   & 583.12   & 1058.43  & -293.37  & -447.76  & 684.97   &&  --      & --       & --       & --       & --       & -- \\
	                        &                 & $Q_{xx}$ & $Q_{yy}$ & $Q_{zz}$ & $Q_{xy}$ & $Q_{xz}$ & $Q_{yz}$ && $Q_{xx}$ & $Q_{yy}$ & $Q_{zz}$ & $Q_{xy}$ & $Q_{xz}$ & $Q_{yz}$ \\\hline
	$^{4}$I$_{15/2}$Z$_{1}$ & This study      & 4.89     & -4.23    & -0.67    & 3.95     & 4.23     & -5.39    && -4.49    & 2.81     & 1.69     & 0.85     & -1.45    & -1.15  \\
	                        & Ref. \cite{horvath}  & 9.32     & -6.37    & -2.95    & 1.92     & 2.26     & -9.55    &&  --      & --       & --       & --       & --       & -- \\
	                        & Ref. \cite{cit:Chen} & 10.40    & -5.95    & -4.44    & -9.12    & -9.96    & -14.32   && -10.5    & -19.5    & 30.0     & -22.8    & -3.1     & -17.7 \\
	$^{4}$I$_{13/2}$Y$_{1}$ & This study      & 5.90     & 0.42     & -6.32    & 2.30     & 4.44     & -5.31    && -4.16    & 3.03     & 1.13     & 0.44     & -2.73    & -0.42 \\
                        	& Ref. \cite{horvath}  & 6.84     & 0.30     & -7.13    & 3.62     & 5.54     & -7.13    &&  --      & --       & --       & --       & --       & -- \\
	\end{tabular}
	\end{ruledtabular}
	\label{tab:SH_params}
\end{table*}

\begin{figure}[h!]
\centering
	\includegraphics[width=0.5\textwidth]{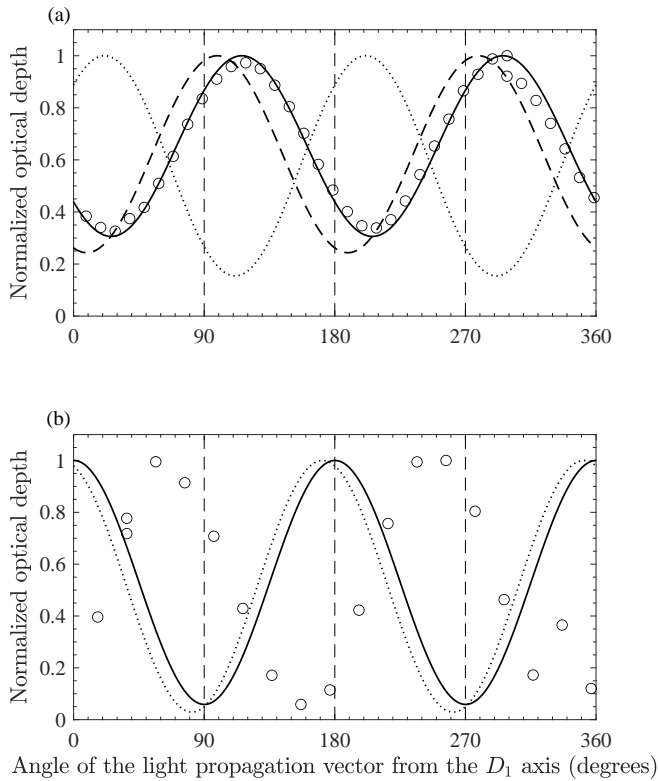}
	\caption{\label{pol} Polarization behavior of the  $^{4}$I$_{15/2}$Z$_{1} \longrightarrow ^{4}$I$_{13/2}$Y$_{1}$ transition for (a) site 1 and (b) site 2. The open circles represent the experimentally determined values \cite{cit:petit}. The solid line is the prediction from this study, the dotted line is the prediction using the crystal-field parameters from \cite{cit:Horvath_thesis} whilst the dashed line is the prediction using the crystal-field parameters from \cite{horvath}.}
\end{figure}
\begin{table}
	\caption{Calculated and experimental $^{4}$I$_{15/2}$Z$_{1}$ ground and $^{4}$I$_{13/2}$Y$_{1}$ excited state hyperfine splittings for site 2 with a magnetic field of 7 T applied along the $D_{1}$ axis \cite{Ran_i__2017}. The calculated splittings were determined from our crystal-field model. All values are in MHz.} 
	\begin{ruledtabular}
	\begin{tabular}{ cccccc }
	 Splittings &   \multicolumn{2}{c}{$^{4}$I$_{15/2}$Z$_{1}$} && \multicolumn{2}{c}{$^{4}$I$_{13/2}$Y$_{1}$} \\\cline{2-3}\cline{5-6} 
	$\Delta E$    & This study & Ref. \cite{Ran_i__2017} &&  This study & Ref. \cite{Ran_i__2017} \\\hline
	$\Delta(1,2)$      & 897 & 995 && 928 & 994 \\
	$\Delta(2,3)$      & 881 & 943 && 912 & 972 \\
	$\Delta(3,4)$      & 865 & 898 && 895 & 953 \\
	$\Delta(4,5)$      & 849 & 862 && 879 & 935 \\
	$\Delta(5,6)$      & 833 & 831 && 863 & 918 \\
	$\Delta(6,7)$      & 817 & 810 && 847 & 903 \\
	$\Delta(7,8)$      & 801 & 796 && 831 & 889 \\
	\end{tabular}
	\end{ruledtabular}
	\label{tab:7T_splittings}
\end{table}

Table \ref{tab:Er_Parameters} shows the fitted free-ion, crystal field, and hyperfine parameters for both sites in Er$^{3+}$:Y$_{2}$SiO$_{5}$. The values obtained by Horvath \textit{et al.} \cite{horvath, cit:Horvath_thesis} are provided for comparison. The two-  and  three-body  interactions and higher order effects in the free-ion Hamiltonian were held fixed to the values obtained by Carnall \textit{et al.} in Er$^{3+}$:LaF$_{3}$ and are given in Table \ref{tab:Fixed_Parameters} \cite{cit:carnall}. The uncertainties of the fitted parameters were estimated through the use of the MCMC techniques in order to sample the posterior probability distribution \cite{cit:aster}. A total of 3 million trials were undertaken for both sites with 343 158 accepted steps for site 1 and 278 705 accepted steps for site 2. This aligns with the Metropolis algorithms $\sim$10 \% acceptance rate recommended for this technique which was fine tuned through altering the step size in the optimization routine \cite{cit:aster}. The algorithm was allowed to `burn in’ and every  10$^{\text{th}}$ element of the last 30 000 steps was selected to ensure that the samples were not correlated.

The calculated energy levels and $g$ values up to the $^{4}$G$_{11/2}$ multiplet of Er$^{3+}$:Y$_{2}$SiO$_{5}$ for site 1 and site 2 are given in Tables \ref{tab:site1} and \ref{tab:site2} respectively, along with the corresponding experimental values. The overall agreement is very good, having standard deviations comparable to the previous work \cite{horvath, cit:Horvath_thesis}, but for considerably more data, distributed over a larger portion of the 4f$^{11}$ configuration.

Spin Hamiltonian parameters derived from the crystal-field model for the $^{4}$I$_{15/2}$Z$_{1}$ and $^{4}$I$_{13/2}$Y$_{1}$ states relevant to the 1.5 $\mu$m telecommunications transition are presented in Table \ref{tab:SH_params}. Those calculated by Horvath \textit{et al.} \cite{horvath}, and  the experimentally determined values of Chen \textit{et al.} and Sun \textit{et al.} \cite{cit:Chen, cit:Sun},  are also included for comparison. The close agreement between our calculated parameters and those derived from measurements indicate that the hyperfine structure of the $^{4}$I$_{15/2}$Z$_{1}$ and $^{4}$I$_{13/2}$Y$_{1}$ states are well accounted for by our fit.

Magnetic-splitting data is crucial to fitting the crystal-field parameters in C$_1$ symmetry. In the absence of magnetic data, an arbitrary rotation about \emph{any} axis leaves the electronic energy levels invariant. Note, however, that there are two magnetically-inequivalent sites, related by a C$_2$ rotation about the $b$ axis of the crystal (our $z$ axis). We list only one set, since the parameter sets for the other orientation may be obtained by multiplying all of the crystal-field parameters with odd $q$ by $-1$. 

A measure of the magnitude of the crystal-field is given by 
crystal-field strength parameters defined as \cite{cit:yeung}:
\begin{equation}
S^{k} = \Bigg[ \frac{1}{2k+1}\Bigg((B^{k}_{0})^2 + 2 \sum_{q>0}\big|B_{q}^{k}\big|^{2}\Bigg)\Bigg]^{1/2}
\label{eq:cf_strength}
\end{equation}
These parameters are listed in Table \ref{tab:Er_Parameters}.  As expected, the crystal-field strength parameters for site 1, identified as the 6-coordinate site, are higher than for the 7-coordinate site 2, particularly the $S^4$ parameter. 

The crystal-field parameters determined in this study differ from previous work by Horvath \textit{et al.} \cite{horvath, cit:Horvath_thesis}. This is to be expected, as we have  added a significant quantity of new data, in the form of electronic energy levels and $g$ values. In previous work, magnetic data were confined to the  $^{4}$I$_{15/2}$Z$_{1}$ and $^{4}$I$_{13/2}$Y$_{1}$ states.


We have demonstrated excellent fits to experimental data from both sites over a wide range of energies. We now address the {\it predictive} ability of the model.  We begin by discussing the optical polarization behavior of the $^{4}$I$_{15/2}$Z$_{1} \rightarrow {^{4}}$I$_{13/2}$Y$_{1}$ transition. We will then discuss high-field Zeeman-hyperfine experiments. 

The precise orientation of transition dipole moments is significant for some applications \cite{Chen592}. We do not yet have a model for electric-dipole moments for rare-earth ions in this crystal, but we can calculate the magnetic dipole moments from the wave functions.   Petit \textit{et al.} \cite{cit:petit} have recently made novel measurements that separate the variation of  magnetic-dipole moments from  the variation of electric-dipole moments of the $^{4}$I$_{15/2}$Z$_{1} \rightarrow {^{4}}$I$_{13/2}$Y$_{1}$ transition for both sites by rotating a cylindrical crystal about the $b$ crystal axis. 

In Fig. \ref{pol}. we show the experimental and calculated angular variations  of the absorption depth of the $^{4}$I$_{15/2}$Z$_{1} \longrightarrow ^{4}$I$_{13/2}$Y$_{1}$ transition as the crystal is rotated about the $b$ axis.  The electric vector  of the light was aligned with the $b$ axis and the propagation vector, $\vec{k}$, was rotated in the $D_{1}$, $D_{2}$ plane (with $\theta = 0$ corresponding to $\vec{k} \parallel D_{1}$, and $\theta = 90^{\circ}$ corresponding to $\vec{k} \parallel D_{2}$). In this orientation, the magnetic vector is also in the $D_{1}$, $D_{2}$ plane. The electric dipole contribution to the absorption depth is constant as the crystal is rotated. 

Our prediction is shown as the solid line in Fig.\ \ref{pol}. We predict the maximum optical depth to be at 295.9$^{\circ}$ for site 1, while the experimental maximum is at 298.6$^{\circ}$.  For site 2, we predict the maximum optical depth to be at 180.1$^{\circ}$, while the experimental maximum is at 256.6$^{\circ}$  \cite{cit:petit}. Thus, the calculation is roughly out of phase with the experimental measurement for site 2.  

Figure \ref{pol} also shows predictions using crystal-field parameters from Horvath \textit{et al.} \cite{horvath, cit:Horvath_thesis}. For site 1, the parameters from Ref.\ \cite{horvath} gives predictions  nearly in phase with the measurements,  while the parameters from Ref.\ \cite{cit:Horvath_thesis} gave predictions that are out of phase with the measurements. The fit in  Ref.\ \cite{cit:Horvath_thesis} did not include data for the hyperfine splitting of the excited state, $^{4}$I$_{13/2}$Y$_{1}$.  Excited state hyperfine data for site 1 is included in Ref.\ \cite{horvath} and our current fit, and this may explain the differences. 
For site 2, neither Ref.\ \cite{cit:Horvath_thesis}, nor the current fit include hyperfine data for the excited state, and it is possible that the addition of such data might bring the calculations into agreement with the measurements. 

We now discuss high-field Zeeman measurements. 
For an isolated Kramers doublet, a spin-Hamiltonian approach breaks down at high magnetic fields, when mixing of electronic states by the Zeeman interaction becomes significant. This non-linearity is apparent even at 4\,T in the data presented in Fig.\ \ref{fig:ErZeeman_site2}.  Ran\v{c}i\'{c} \textit{et al.}\ \cite{Ran_i__2017} have measured hyperfine splittings for the $^{4}$I$_{15/2}$Z$_{1}$ and excited $^{4}$I$_{13/2}$Y$_{1}$ states at a magnetic field of 7 T applied along the $D_{1}$ axis for site 2. 
Table \ref{tab:7T_splittings} compares predictions from our crystal-field model with the  experimental splittings.  The predictions  agree with the experimentally determined values to within 10 \%.
The site 2 fit does not contain any data for the hyperfine splitting of the excited state, and with more data in the fit, better agreement could be expected.

\section{Conclusions}

We have reported a parametrized crystal-field analysis for both C$_{1}$ symmetry sites in Er$^{3+}$ doped Y$_{2}$SiO$_{5}$ which uses crystal-field energy levels up to 27 000 cm$^{-1}$, approximately 70 electronic g-values per site and electron-nuclear interactions within crystal-field levels of the two lowest energy multiplets as data inputs. The crucial feature of the analysis performed here is that the data input includes directional information (i.e. $g$ values) which spans a considerable portion of the entire 4f$^{11}$ configuration; this is required to obtain well determined crystal-field parameters in sites of such low point group symmetry. We demonstrate that the analysis presented here can {\it predict} the high-field Zeeman-hyperfine splittings (splittings measured in the non-linear regime \cite{Ran_i__2017}) as well as optical polarization behavior for the telecommunications transitions near 1.5 $\mu m$ \cite{cit:petit}.

\begin{acknowledgments}
N.L.J. would like to thank the Dodd-Walls Centre for Photonic and Quantum Technologies for the provision of a PhD studentship. The technical assistance of Mr. S. Hemmingson, Mr. R. J. Thirkettle and Mr. G. MacDonald is gratefully acknowledged.
\end{acknowledgments}

\end{document}